\documentclass[aps,pra,twocolumn,reprint]{revtex4-2}
\usepackage{amsmath,amssymb}
\usepackage{hyperref}
\usepackage{graphicx}

\begin{document}

\title{Growth versus equity in a stochastic multiplicative wealth model:\\ The cost of redistribution and the emergence of an inequality Laffer curve}

\author{Renato Vieira dos Santos}
\email{renato.santos@ufla.br}
\affiliation{Instituto de Ci\^encia, Tecnologia e Inova\c{c}\~ao -- ICTIN, Universidade Federal de Lavras -- UFLA, Campus Para\'iso, MG 37950-000, Brazil}

\begin{abstract}
A stochastic multiplicative wealth model with taxation and redistribution is extended to include two channels through which fiscal policy can reduce the economy's average growth rate: productive disincentive (a tax‑dependent mean return) and administrative inefficiency (loss of part of the collected revenue). For uniform redistribution, the stationary density of normalized wealth remains an inverse‑gamma law, but its shape parameter acquires a richer dependence on the tax rate. The central finding is the emergence of an \emph{inequality Laffer curve}: when the disincentive is sufficiently strong, the Gini coefficient first decreases, reaches a minimum at a finite tax rate, and then increases again as taxation intensifies. In the extreme limit the stationary state collapses back to the condensed phase, restoring the oligarchic transition. Both channels also create a direct trade‑off between inequality and long‑run average wealth, giving rise to a welfare‑maximizing tax rate. The two channels are combined into a single analytically tractable model, and all predictions are validated by agent‑based simulations. The results show that the qualitative existence of an optimal tax rate is robust whenever the aggregate growth rate responds to fiscal policy, while the inequality Laffer curve is a specific prediction of the productive‑disincentive channel.
\end{abstract}


\maketitle

\section{Introduction}
\label{sec:intro}

The study of wealth and income distributions has a long and distinguished history, dating back to the seminal observation by Pareto that the upper tail of the distribution follows a power law \cite{Pareto1896}. This empirical regularity has been confirmed across countries and epochs \cite{Mandelbrot1963,Champernowne1953,Atkinson1970b,Piketty2014}, and has motivated a vast theoretical literature aimed at uncovering the microscopic mechanisms that generate it. Among the simplest and most successful explanations is the multiplicative stochastic process, in which individual wealth evolves through successive random proportional shocks \cite{Gibrat1931}. When left unchecked, such processes produce lognormal distributions with ever-increasing variance, eventually concentrating almost all wealth in the hands of a vanishingly small fraction of the population \cite{Bouchaud2000,Levy1996}. More recent theoretical refinements have shown how the interplay between idiosyncratic risk and heterogeneous returns can robustly generate Pareto tails \cite{Gabaix2009}.

Real economies, however, rarely exhibit this extreme condensation. Fiscal institutions (taxation and redistribution) act as mean-reverting forces that counteract the multiplicative amplification and stabilize the wealth distribution. A particularly elegant formulation of this interplay was proposed by de Oliveira \cite{deOliveira2017,deOliveira2020}, who considered a population of agents subject to idiosyncratic multiplicative shocks, followed by a tax on the realized wealth and a subsequent redistribution of the collected revenue. The model possesses a rich phase diagram, including a critical point at which condensation occurs in the absence of redistribution and a non-condensed phase when redistribution is active. Recent work by Barros, Martins, and Anteneodo \cite{Barros2026} generalized the redistribution protocol to a broad class of state-dependent transfer mechanisms and developed a Fokker--Planck description that yields exact stationary distributions for several important cases, including uniform redistribution and conditional cash transfer programs.

The models mentioned above share a common feature: the average growth factor of wealth, denoted by $\nu$, is treated as an exogenous constant. As a consequence, the aggregate wealth of the population grows at a rate that is entirely independent of the tax rate $A$ or the details of the redistribution rule. Fiscal policy affects only how the pie is sliced, not how fast the pie grows. This separation between distribution and growth is analytically convenient, and it has allowed the derivation of closed-form stationary distributions and Gini coefficients \cite{Barros2026,Lima2022,Boghosian2017}. From the standpoint of economic realism, however, the assumption is restrictive. A large body of theoretical and empirical work in public economics suggests that taxation influences labor supply, saving behavior, and investment decisions, thereby affecting the growth rate of the economy \cite{Atkinson1970,Saez2001,Piketty2014,Mankiw2009}. In the econophysics literature, several authors have addressed this issue by introducing tax-dependent returns or by allowing agents to choose between risky and risk-free assets whose returns are affected by fiscal policy \cite{During2009,Bassetti2011,During2008,Bertotti2012,Toscani2014}. The broader landscape of kinetic wealth models has been comprehensively reviewed by Yakovenko and Rosser \cite{Yakovenko2009}.

The present paper builds on the analytical framework of Ref.\ \cite{Barros2026} and extends it by incorporating two minimal channels through which redistribution can depress the average wealth of the economy. The first channel, which we call \emph{productive disincentive}, supposes that the mean return on wealth decreases linearly with the tax rate, $\nu = \nu_0 - \gamma A$, with $\gamma \ge 0$. This is the simplest functional form that captures the idea that taxation reduces the net reward to productive activity. The second channel, termed \emph{administrative inefficiency}, assumes that a fraction $\delta \in [0,1)$ of the tax revenue is lost during the collection and redistribution process, so that only a fraction $1-\delta$ of the collected amount actually reaches the agents. This reflects the real-world costs of bureaucracy, means-testing, and rent-seeking that accompany any fiscal system \cite{Okun1975}. Both modifications preserve the analytical tractability of the uniform-redistribution benchmark: the stationary density remains an inverse-gamma law, but its shape parameter acquires a richer dependence on the tax rate.

The central findings of this work are twofold. First, we uncover a dramatic non-monotonic regime in the productive-disincentive model. When the sensitivity $\gamma$ exceeds a critical threshold, the stationary Gini coefficient does not decrease monotonically with $A$, but instead reaches a minimum at a finite tax rate and then \emph{increases} as taxation intensifies. This ``inequality Laffer curve'' demonstrates that beyond an optimal rate, further taxation aggravates wealth concentration instead of reducing it. In the extreme limit $\gamma A \to \nu_0$, the stationary state collapses back to the condensed phase, restoring the oligarchic transition. Second, both mechanisms generate a genuine trade-off between equity and aggregate output. As $A$ increases, the stationary Gini coefficient generally falls, but the long-run average wealth after a fixed horizon also falls. When these two objectives are combined into a social welfare function, such as the Atkinson index \cite{Atkinson1970}, an interior optimal tax rate $A^*$ appears for any positive degree of inequality aversion. We derive closed-form expressions for $A^*$ in the uniform-redistribution case and show how it responds to the parameters $\gamma$ and $\delta$. The two channels are then combined into a single model, and the interplay between them is mapped numerically. Agent-based simulations are used throughout to validate the theoretical predictions.

The remainder of the paper is organized as follows. Section \ref{sec:original} recalls the original model of Ref.\ \cite{Barros2026} and its stationary solution for uniform redistribution. Section \ref{sec:model1} introduces the productive-disincentive mechanism, with special emphasis on the non-monotonic regime and the inequality Laffer curve (Sec.~\ref{sec:nonmonotonic}). Section \ref{sec:model2} does the same for the administrative-inefficiency channel. The combined model is developed in Sec.\ \ref{sec:model3}, where we also present a consolidated validation of the stationary Gini coefficients across all three models, discuss the aggregate growth trade-off, and map the optimal tax rate in the $(\gamma,\delta)$ plane. Conclusions and directions for future work are offered in Sec.\ \ref{sec:conclusion}.

\section{The original model and its stationary solution}
\label{sec:original}

Consider a population of $N$ agents. At each discrete time step $t$, the wealth of agent $i$ is updated according to a three-stage protocol \cite{deOliveira2017,Barros2026}. First, it is multiplied by a stochastic growth factor $f_{i,t} = \nu + \varepsilon \eta_{i,t}$, where $\nu > 0$ is the mean return, $\varepsilon > 0$ sets the amplitude of the fluctuations, and the $\eta_{i,t}$ are independent standard Gaussian random variables. Second, a fraction $A \in (0,1)$ of the resulting wealth is collected as tax. Third, the total tax revenue is redistributed among the agents according to weights $r_{i,t}$, with $\sum_i r_{i,t} = 1$. The evolution of individual wealth is therefore
\begin{equation}
W_{i,t+1} = (1-A) f_{i,t} W_{i,t} + r_{i,t} A \sum_{j=1}^N f_{j,t} W_{j,t}.
\label{eq:map_original}
\end{equation}
In the original de Oliveira model \cite{deOliveira2017}, the tax rate could depend on the wealth share of each agent, but here we restrict ourselves to the uniform tax rate $A$, which isolates the effects of the redistribution rule.

To make analytical progress, one resorts to a mean-field closure in which the fluctuating tax pool is replaced by its typical value \cite{Barros2026}:
\begin{equation}
\sum_{j=1}^N f_{j,t} W_{j,t} \approx N \nu \langle W \rangle_t,
\label{eq:mf_pool}
\end{equation}
where $\langle W \rangle_t = \frac{1}{N}\sum_j W_{j,t}$ is the mean wealth at time $t$. This approximation becomes exact in the large-$N$ limit when correlations among agents are negligible. Introducing the mean-normalized wealth
\begin{equation}
y_t = \frac{W_t}{\langle W \rangle_t},
\label{eq:y_def}
\end{equation}
which satisfies $\langle y \rangle_t = 1$ by construction, and taking the continuous-time limit in the It\^o convention \cite{Gardiner2009}, one arrives at the stochastic differential equation (SDE) for a representative agent \cite{Barros2026}:
\begin{equation}
dy = A\!\left[ \frac{\phi(y)}{\langle \phi \rangle_t} - y \right] dt + s\, y\, dB_t,
\label{eq:SDE_original}
\end{equation}
The It\^o interpretation is the appropriate one for this limit because the discrete-time process has independent increments and the multiplicative noise acts on the current (non-anticipative) value of wealth. A Stratonovich interpretation would introduce an additional drift term $\frac{1}{2}s^2 y$ that is absent in the original microscopic dynamics and would alter the stationary state. In Eq.\ \eqref{eq:SDE_original}, $B_t$ is a standard Wiener process, $s = \sqrt{2D}\,(1-A)$ with
\begin{equation}
D = \frac{1}{2}\!\left(\frac{\varepsilon}{\nu}\right)^{\!2},
\label{eq:D_def}
\end{equation}
and the redistribution kernel $\phi(y)$ encodes the allocation rule via $r_{i,t} = \phi(y_{i,t}) / \sum_j \phi(y_{j,t})$. The ensemble average $\langle \phi \rangle_t$ is taken with respect to the time-dependent density of $y$.

The associated Fokker--Planck equation for the probability density $P(y,t)$ reads
\begin{equation}
\partial_t P = A\, \partial_y\!\left[ \left(y - \frac{\phi(y)}{\langle \phi \rangle_t}\right) P \right] + \frac{s^2}{2}\, \partial_y^2\!\left( y^2 P \right).
\label{eq:FPE_original}
\end{equation}
In the long-time limit the system reaches a stationary state with vanishing probability current. Setting $\partial_t P = 0$ and integrating once yields the general stationary solution
\begin{equation}
P_{\rm st}(y) = \mathcal{N}\, y^{-2-\alpha} \exp\!\left[ \frac{\alpha}{\langle \phi \rangle_{\rm st}} \int^y \frac{\phi(u)}{u^2}\, du \right],
\label{eq:Pst_general}
\end{equation}
where $\mathcal{N}$ is a normalization constant,
\begin{equation}
\alpha = \frac{2A}{s^2} = \frac{A}{D(1-A)^2},
\label{eq:alpha_original}
\end{equation}
and $\langle \phi \rangle_{\rm st}$ must be determined self-consistently. The existence of a stationary state for any non-regressive kernel $\phi(y)$ with a positive floor at the origin was demonstrated in Ref.\ \cite{Barros2026}.

\subsection{Uniform redistribution benchmark}
\label{sec:uniform}

The simplest redistribution rule assigns an equal share of the tax pool to every agent, corresponding to $\phi(y) \equiv 1$. In this case the self-consistency condition is trivially satisfied because $\langle \phi \rangle_{\rm st} = 1$, and Eq.\ \eqref{eq:Pst_general} reduces to an inverse-gamma distribution,
\begin{equation}
P_{\rm st}(y) = \frac{\alpha^{\alpha+1}}{\Gamma(\alpha+1)}\, y^{-2-\alpha}\, e^{-\alpha/y},
\label{eq:Pst_uniform}
\end{equation}
which depends on the single parameter $\alpha$ defined in Eq.\ \eqref{eq:alpha_original}. The distribution is normalizable for any $\alpha > 0$, but its second moment is finite only when $\alpha > 1$. The stationary Gini coefficient can be computed in closed form \cite{Barros2026} by evaluating the integral
\begin{equation}
G_{\rm st} = 1 - 2\int_0^\infty y\, P_{\rm st}(y)\, S_{\rm st}(y)\, dy,
\label{eq:Gini_def}
\end{equation}
where $S_{\rm st}(y) = \int_y^\infty P_{\rm st}(u)\, du$ is the complementary cumulative distribution. Substituting Eq.\ \eqref{eq:Pst_uniform} and performing the integration gives
\begin{equation}
G_{\rm st} = 1 - 2\, I_{1/2}(\alpha+1,\alpha),
\label{eq:Gini_uniform}
\end{equation}
with $I_x(a,b)$ the regularized incomplete beta function \cite{Abramowitz1964}. Equation \eqref{eq:Gini_uniform} shows that $G_{\rm st}$ is a monotonically decreasing function of $\alpha$, and hence of the tax rate $A$ for fixed $D$.

The aggregate dynamics of the model is governed by the mean wealth. From Eq.\ \eqref{eq:map_original} and the mean-field approximation one finds $\langle W \rangle_{t+1} \approx \nu \langle W \rangle_t$, so that the economy grows at a constant rate $\nu$ regardless of the tax rate $A$. This feature makes the uniform-redistribution model a convenient benchmark, but it also makes it unable to address the question that motivates the present work: what is the cost of redistribution in terms of foregone growth?

\section{Model I: Productive disincentive}
\label{sec:model1}

\subsection{Motivation and modification}
\label{sec:model1_motivation}

The first mechanism we introduce is a tax-dependent average return. We assume that the mean growth factor decreases linearly with the tax rate,
\begin{equation}
\nu = \nu_0 - \gamma A,
\label{eq:nu_A}
\end{equation}
where $\nu_0 > 0$ is the baseline return in the absence of taxation and $\gamma \ge 0$ quantifies the sensitivity of productivity to the tax. Both $\nu_0$ and $\gamma$ are dimensionless constants. For the model to be meaningful we require $\nu_0 - \gamma A > 0$ for all $A$ of interest; this imposes $\gamma < \nu_0/A$, which is easily satisfied for realistic parameter ranges.

The economic motivation behind Eq.\ \eqref{eq:nu_A} is deliberately stylized. A proportional tax on wealth reduces the net return that agents obtain on their assets, which may lower the incentive to engage in productive investment, to supply labor, or to undertake entrepreneurial risk. At the aggregate level this translates into a slower growth of the average wealth. Similar ideas have been explored in kinetic models of wealth distribution with risky investments \cite{Bassetti2011,Toscani2014,During2009}, where the after-tax return affects the optimal portfolio allocation. Equation \eqref{eq:nu_A} is the simplest functional form that captures this feedback while preserving the analytical tractability of the Fokker--Planck description.

\subsection{Modified stochastic differential equation}
\label{sec:model1_sde}

The microscopic dynamics remain governed by Eq.\ \eqref{eq:map_original}, but now with $f_{i,t} = (\nu_0 - \gamma A) + \varepsilon \eta_{i,t}$. The mean-field replacement in Eq.\ \eqref{eq:mf_pool} continues to hold, provided we replace $\nu$ by $\nu(A)$. The derivation of the SDE for $y$ proceeds exactly as in Ref.\ \cite{Barros2026}, with the sole difference that the constants $D$ and $s$ acquire a dependence on $A$ through $\nu(A)$. Specifically,
\begin{equation}
D(A) = \frac{1}{2}\!\left( \frac{\varepsilon}{\nu_0 - \gamma A} \right)^{\!2},
\label{eq:D_A}
\end{equation}
and $s(A) = \sqrt{2D(A)}\,(1-A)$. The effective parameter $\alpha$ becomes
\begin{equation}
\alpha_{\rm I}(A) = \frac{A}{D(A)(1-A)^2} = \frac{2A}{\varepsilon^2}\, (1-A)^{-2}\, (\nu_0 - \gamma A)^2.
\label{eq:alpha_I}
\end{equation}
For $\gamma=0$, Eq.\ \eqref{eq:alpha_I} reduces to the original expression \eqref{eq:alpha_original}. The SDE for a generic redistribution kernel retains the form \eqref{eq:SDE_original}, with $s$ and $\langle \phi \rangle_t$ now evaluated using the modified $D(A)$. The Fokker--Planck equation \eqref{eq:FPE_original} and its stationary solution \eqref{eq:Pst_general} remain formally unchanged; only the numerical value of $\alpha$ is altered.

\subsection{Stationary distribution for uniform redistribution}
\label{sec:model1_stationary}

When redistribution is uniform, $\phi(y)=1$, the stationary FPE still yields the inverse-gamma density \eqref{eq:Pst_uniform}, but with $\alpha$ given by $\alpha_{\rm I}(A)$ instead of Eq.\ \eqref{eq:alpha_original}. All statistical properties of the stationary state therefore follow from those of the inverse-gamma law with shape parameter $\alpha_{\rm I}(A)$. In particular, the Gini coefficient is
\begin{equation}
G_{\rm st}^{\rm I}(A) = 1 - 2\, I_{1/2}\bigl(\alpha_{\rm I}(A)+1,\alpha_{\rm I}(A)\bigr),
\label{eq:Gini_I}
\end{equation}
and the stationary variance is $\operatorname{Var}[y] = 1/(\alpha_{\rm I}(A)-1)$ for $\alpha_{\rm I}(A) > 1$.

\subsection{The non-monotonic regime: an inequality Laffer curve}
\label{sec:nonmonotonic}

The most striking consequence of Eq.\ \eqref{eq:alpha_I} is the possibility of a non-monotonic dependence of $\alpha_{\rm I}(A)$ on $A$, and therefore of the Gini coefficient. Differentiating $\alpha_{\rm I}(A)$ with respect to $A$ yields
\begin{equation}
\frac{d\alpha_{\rm I}}{dA} = \frac{2}{\varepsilon^2} \frac{(\nu_0 - \gamma A)}{(1-A)^3}
\bigl[ (\nu_0 - \gamma A)(1+A) - 2\gamma A(1-A) \bigr].
\label{eq:dalpha}
\end{equation}
The sign of $d\alpha_{\rm I}/dA$ is controlled by the expression in square brackets. Setting it to zero gives the condition for an extremum,
\begin{equation}
(\nu_0 - \gamma A)(1+A) = 2\gamma A(1-A),
\end{equation}
which simplifies to the quadratic equation
\begin{equation}
\gamma A^2 + (\nu_0 - 3\gamma)A + \nu_0 = 0.
\label{eq:quadratic}
\end{equation}
For small $\gamma$, Eq.\ \eqref{eq:quadratic} has no real root inside the admissible interval $(0,\nu_0/\gamma)$, and $\alpha_{\rm I}$ increases monotonically. When $\gamma$ exceeds a critical value, the discriminant $\Delta = \nu_0^2 - 10\nu_0\gamma + 9\gamma^2$ becomes positive and a root $A_c$ appears in the admissible interval. For $A < A_c$, $d\alpha_{\rm I}/dA > 0$; for $A > A_c$, $d\alpha_{\rm I}/dA < 0$. Consequently, $\alpha_{\rm I}$ attains a maximum at $A_c$, and the Gini coefficient, which is a monotonically decreasing function of $\alpha$, attains a minimum. This non-monotonic behavior is the \emph{inequality Laffer curve}.

\begin{figure*}[!htb]
\includegraphics[width=\textwidth]{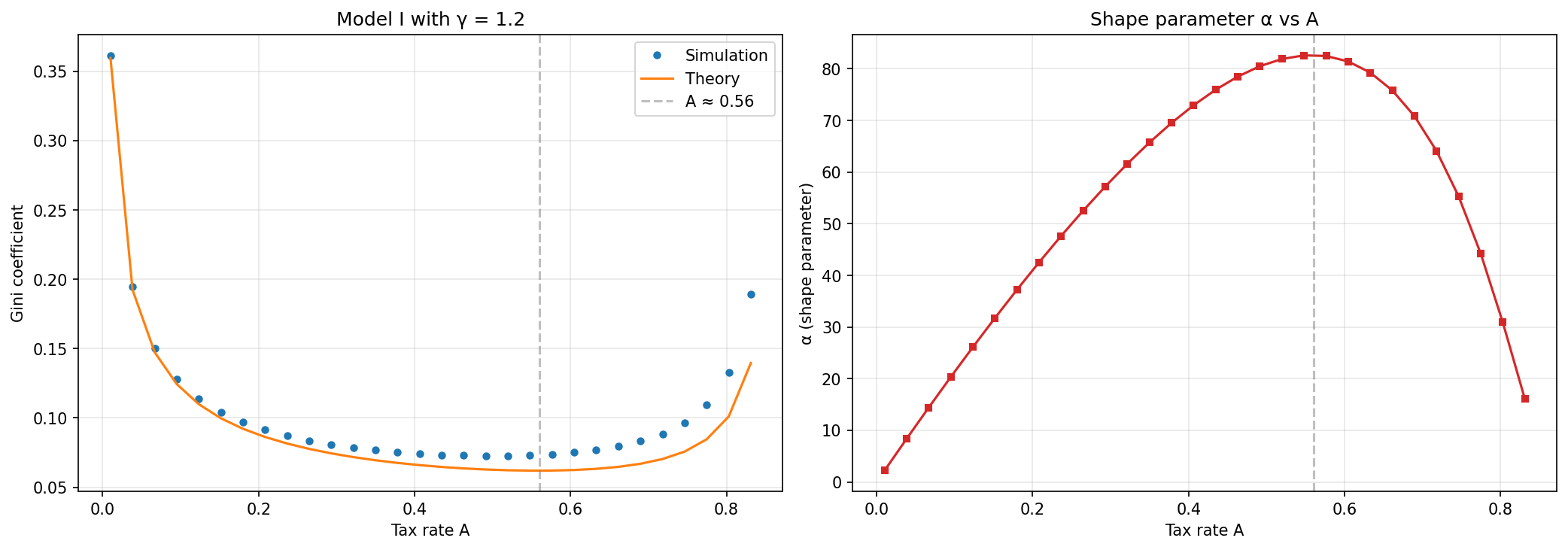}
\caption{Stationary Gini coefficient (left panel) and shape parameter $\alpha$ (right panel) as functions of the tax rate $A$ for $\gamma=1.2$ in Model~I. Symbols represent agent-based simulations ($N=10^4$, $\nu_0=1.05$, $\varepsilon=0.1$, $10^4$ time steps) and solid lines are the analytical predictions of Eqs.~\eqref{eq:Gini_I} and \eqref{eq:alpha_I}. The Gini reaches a minimum near $A=0.56$ and then increases, revealing the inequality Laffer curve. The shape parameter exhibits a maximum at the same point, confirming the non-monotonicity.}
\label{fig:Gini_I_gamma12}
\end{figure*}

For the parameters used in our simulations ($\nu_0=1.05$, $\varepsilon=0.1$), the critical value of $\gamma$ above which the non-monotonic regime appears is $\gamma_c \approx 0.37$. Figure~\ref{fig:Gini_I_gamma12} displays the phenomenon for $\gamma=1.2$, well above the threshold. The Gini coefficient decreases until $A \approx 0.56$, reaches a minimum value of about $0.20$, and then increases again. The right panel shows the corresponding behavior of $\alpha_{\rm I}$: it reaches a maximum at the same point, after which the destabilizing effect of the depressed mean return overwhelms the mean-reverting drift from redistribution.

In the extreme limit $\gamma A \to \nu_0$, Eq.\ \eqref{eq:alpha_I} gives $\alpha_{\rm I} \to 0$. The inverse-gamma distribution ceases to be normalizable, and the system returns to the condensed phase where a single agent owns the entire wealth. Thus, excessive taxation not only fails to reduce inequality but restores the oligarchic transition that redistribution was designed to prevent. This is a dramatic counterexample to the intuition that higher taxes always reduce inequality, and it is one of the main contributions of the present work.

To the best of our knowledge, this non-monotonic behavior of the Gini coefficient with respect to the tax rate has not been reported before in the context of exactly solvable multiplicative wealth models. The literature on kinetic wealth equations has largely focused on mechanisms that preserve a monotonic relationship between taxation and equality \cite{During2009,Bassetti2011,Barros2026}, or that introduce inefficiencies which reduce the equalizing power of redistribution without reversing it \cite{Bertotti2012}. The present result shows that a reversal is possible even within the simplest extension of the de Oliveira framework, provided the feedback of taxation on the productive capacity of the economy is taken into account. The analytical transparency of the model allows us to identify precisely the condition $\gamma > \gamma_c$ under which the reversal occurs, and to connect the extreme limit of this regime with the known oligarchic transition of the original model. This connection, we believe, adds a new layer of understanding to the phase diagram of wealth condensation models.

\subsection{Growth of aggregate wealth}
\label{sec:model1_growth}

The crucial difference from the original model lies in the dynamics of the mean wealth. From the microscopic rule \eqref{eq:map_original} and the mean-field approximation, one finds
\begin{equation}
\langle W \rangle_{t+1} \approx \nu(A) \langle W \rangle_t = (\nu_0 - \gamma A) \langle W \rangle_t.
\label{eq:mean_growth_I}
\end{equation}
After $T$ time steps, starting from an initial mean $W_0$, the expected aggregate wealth is
\begin{equation}
\langle W \rangle_T \approx W_0\, (\nu_0 - \gamma A)^T.
\label{eq:aggregate_T}
\end{equation}
Thus, even though the distribution of $y = W/\langle W \rangle$ reaches a stationary state, the absolute scale of the economy grows at a rate that declines linearly with $A$. A larger tax rate reduces inequality (at least up to the turning point) but also lowers the long-run level of average wealth.

\subsection{The trade-off and optimal taxation}
\label{sec:model1_tradeoff}

Equation \eqref{eq:aggregate_T} and Eq.\ \eqref{eq:Gini_I} together define a trade-off: increasing $A$ lowers $G_{\rm st}^{\rm I}$ (good for equity) but also lowers $\langle W \rangle_T$ (bad for aggregate output). To make this trade-off operational we need a social welfare function that combines both dimensions. A standard choice is the Atkinson welfare index \cite{Atkinson1970}
\begin{equation}
\mathcal{W}_\epsilon = \langle W \rangle \left( \frac{1}{N} \sum_i y_i^{1-\epsilon} \right)^{\!1/(1-\epsilon)},
\label{eq:Atkinson}
\end{equation}
where $\epsilon \ge 0$ measures the degree of inequality aversion. When $\epsilon = 0$, $\mathcal{W}_0 = \langle W \rangle$ and the planner cares only about total output. As $\epsilon \to \infty$, the index approaches the Rawlsian criterion $\mathcal{W}_\infty = \min_i W_i$ \cite{Rawls1971}. For $\epsilon = 1$, Eq.\ \eqref{eq:Atkinson} reduces to the Nash--Bernoulli form $\mathcal{W}_1 = \langle W \rangle \exp(\langle \ln y \rangle)$.

In the stationary state, the expectation $\langle y^{1-\epsilon} \rangle$ can be computed analytically for the inverse-gamma density. Using Eq.\ \eqref{eq:Pst_uniform} we obtain
\begin{equation}
\langle y^{1-\epsilon} \rangle = \frac{\Gamma(\alpha + \epsilon)}{\Gamma(\alpha+1)}\, \alpha^{\epsilon-1},
\label{eq:moment_gamma}
\end{equation}
valid for $\alpha > 1-\epsilon$. Substituting into Eq.\ \eqref{eq:Atkinson} and using Eq.\ \eqref{eq:aggregate_T} for the mean wealth after a fixed horizon $T$, the welfare index becomes a known function of $A$, parameterized by $\nu_0$, $\gamma$, $\varepsilon$, $\epsilon$, and $T$.

The behavior of $\mathcal{W}_\epsilon(A)$ is instructive. For $\epsilon = 0$, the welfare function reduces to $\langle W \rangle_T$, which decreases linearly with $A$, and the optimal policy is $A^* = 0$: in the absence of inequality aversion, any tax that reduces growth is welfare-destroying. For any $\epsilon > 0$, however, the welfare function acquires an additional factor that increases with $A$ (because higher taxes reduce inequality and raise the Atkinson adjustment). As a result, $\mathcal{W}_\epsilon(A)$ exhibits a unique interior maximum at some tax rate $A^*_{\rm I}(\gamma,\epsilon)$. This optimal rate increases with $\epsilon$ (more inequality-averse societies prefer more redistribution) and decreases with $\gamma$ (when taxation is more damaging to growth, the optimal rate is lower). In the non-monotonic regime, the optimal rate is further constrained by the fact that beyond the minimum of the Gini coefficient, both equity and output deteriorate simultaneously, making a strong case for moderate taxation.

\section{Model II: Administrative inefficiency}
\label{sec:model2}

\subsection{Motivation and modification}
\label{sec:model2_motivation}

The second channel through which redistribution can depress aggregate wealth is the loss of resources during the fiscal process. Real tax systems incur administrative costs, and means-tested transfer programs require bureaucracy that consumes part of the collected revenue \cite{Okun1975,Bertotti2012,Calvelli2023}. We model this by assuming that only a fraction $1-\delta$ of the tax pool actually reaches the agents, with $\delta \in [0,1)$. The microscopic evolution now reads
\begin{equation}
W_{i,t+1} = (1-A) f_{i,t} W_{i,t} + r_{i,t}\, (1-\delta) A \sum_{j=1}^N f_{j,t} W_{j,t}.
\label{eq:map_delta}
\end{equation}
All other elements of the model (the multiplicative shocks, the uniform tax rate, and the redistribution weights) remain unchanged.

\subsection{Effective SDE and the modified parameter}
\label{sec:model2_sde}

Applying the mean-field approximation \eqref{eq:mf_pool} to Eq.\ \eqref{eq:map_delta} and passing to continuous time leads to the SDE for $y$:
\begin{equation}
dy = A\!\left[ (1-\delta) \frac{\phi(y)}{\langle \phi \rangle_t} - y \right] dt + s\, y\, dB_t,
\label{eq:SDE_delta}
\end{equation}
with $s = \sqrt{2D}\,(1-A)$ and $D$ given by the original expression \eqref{eq:D_def} (since $\nu$ is unaffected by $\delta$). The only change relative to Eq.\ \eqref{eq:SDE_original} is the prefactor $(1-\delta)$ multiplying the redistribution term.

For uniform redistribution, $\phi(y)=1$ and $\langle \phi \rangle_t = 1$. The SDE simplifies to
\begin{equation}
dy = A\bigl[ (1-\delta) - y \bigr] dt + s\, y\, dB_t,
\label{eq:SDE_delta_uniform}
\end{equation}
which is again a mean-reverting process with linear drift. The associated stationary FPE can be solved exactly. The stationary density is still an inverse-gamma law,
\begin{equation}
P_{\rm st}^{\rm II}(y) = \frac{\alpha_{\rm II}^{\alpha_{\rm II}+1}}{\Gamma(\alpha_{\rm II}+1)}\, y^{-2-\alpha_{\rm II}}\, e^{-\alpha_{\rm II}/y},
\label{eq:Pst_delta}
\end{equation}
but now the shape parameter is
\begin{equation}
\alpha_{\rm II}(A) = \frac{A}{D(1-A)^2}\, (1-\delta).
\label{eq:alpha_II}
\end{equation}
The only difference from the original $\alpha$ is the factor $(1-\delta)$, which reduces the effective strength of the mean-reverting drift. A less efficient redistribution system (larger $\delta$) leads to a smaller $\alpha_{\rm II}$ and therefore to higher stationary inequality. The Gini coefficient is given by Eq.\ \eqref{eq:Gini_uniform} with $\alpha \to \alpha_{\rm II}(A)$.

\subsection{Aggregate growth and the trade-off}
\label{sec:model2_growth}

From Eq.\ \eqref{eq:map_delta}, the evolution of mean wealth follows
\begin{equation}
\langle W \rangle_{t+1} \approx \nu\, \langle W \rangle_t - \delta A\, \nu\, \langle W \rangle_t = \nu (1 - \delta A) \langle W \rangle_t.
\label{eq:mean_growth_II}
\end{equation}
The effective growth rate per period is $\nu (1 - \delta A)$, which decreases linearly with $A$. After $T$ periods, the expected aggregate wealth is $\langle W \rangle_T \approx W_0\, [\nu(1-\delta A)]^T$. As in Model~I, a higher tax rate reduces the long-run scale of the economy, but through a different mechanism: here the loss arises because resources are destroyed in the redistribution pipeline rather than because productivity is discouraged. The two channels have similar aggregate consequences, yet they affect the stationary distribution in distinct ways. In Model~I, $\alpha_{\rm I}(A)$ contains a factor $(\nu_0 - \gamma A)^2$ that can make it non-monotonic in $A$, whereas in Model~II, $\alpha_{\rm II}(A)$ is simply proportional to $A$, as in the original model, but scaled down by $(1-\delta)$. No inequality Laffer curve appears in Model~II.

The Atkinson welfare index for Model~II can be computed in closed form using Eqs.\ \eqref{eq:Atkinson}, \eqref{eq:moment_gamma}, and $\alpha_{\rm II}(A)$. The resulting optimal tax rate $A^*_{\rm II}(\delta,\epsilon)$ decreases with $\delta$: societies with less efficient bureaucracies should tax less, because a larger fraction of the collected revenue would be wasted before reaching the intended beneficiaries. Agent-based simulations confirm the analytical predictions for the stationary distribution and the Gini coefficient across the relevant parameter range.

\section{Model III: Combined effects}
\label{sec:model3}

\subsection{Formulation and stationary solution}
\label{sec:model3_formulation}

We now bring together the two channels analyzed separately. The average return becomes $\nu = \nu_0 - \gamma A$, and a fraction $\delta$ of the tax revenue is lost. The microscopic evolution is
\begin{equation}
\begin{split}
W_{i,t+1} &= (1-A) \bigl[ (\nu_0 - \gamma A) + \varepsilon \eta_{i,t} \bigr] W_{i,t}\\& + r_{i,t}\, (1-\delta) A \sum_{j=1}^N f_{j,t} W_{j,t}.
\label{eq:map_combined}
\end{split}
\end{equation}
The mean-field SDE for uniform redistribution takes the compact form
\begin{equation}
dy = A\bigl[ (1-\delta) - y \bigr] dt + s(A)\, y\, dB_t,
\label{eq:SDE_combined}
\end{equation}
with $s(A) = \sqrt{2D(A)}\,(1-A)$ and $D(A)$ given by Eq.\ \eqref{eq:D_A}. The stationary density remains inverse-gamma, with shape parameter
\begin{equation}
\begin{split}
\alpha_{\rm III}(A) &= \frac{A}{D(A)(1-A)^2}\, (1-\delta) \\&= \frac{2A}{\varepsilon^2}\, (1-A)^{-2}\, (\nu_0 - \gamma A)^2\, (1-\delta).
\label{eq:alpha_III}
\end{split}
\end{equation}
Equation \eqref{eq:alpha_III} interpolates smoothly between Model~I ($\delta=0$) and Model~II ($\gamma=0$). The Gini coefficient and the aggregate growth factor become, respectively,
\begin{align}
G_{\rm st}^{\rm III}(A) &= 1 - 2\, I_{1/2}\bigl(\alpha_{\rm III}(A)+1, \alpha_{\rm III}(A)\bigr), \label{eq:Gini_III} \\
\langle W \rangle_{T} &\approx W_0\, \bigl[ (\nu_0 - \gamma A)(1 - \delta A) \bigr]^T. \label{eq:growth_III}
\end{align}
In the combined model, the condition for the inequality Laffer curve is modified by the factor $(1-\delta)$, but the qualitative non-monotonicity persists for sufficiently large $\gamma$.

\subsection{Validation of the stationary distributions and shape parameters}
\label{sec:validation}

\begin{figure*}[!htb]
\includegraphics[width=\textwidth]{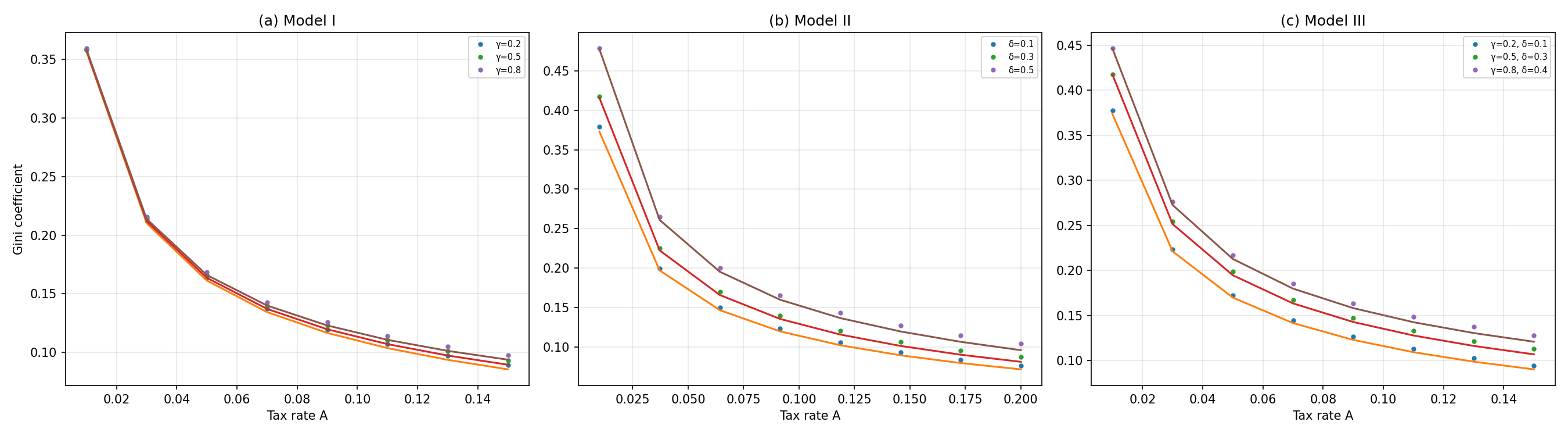}
\caption{Stationary Gini coefficient as a function of the tax rate $A$ for moderate parameter values. (a) Model~I for three values of $\gamma$. (b) Model~II for three values of $\delta$. (c) Model~III for three representative pairs $(\gamma,\delta)$. Symbols correspond to agent-based simulations and solid lines to the theoretical expressions \eqref{eq:Gini_I}, \eqref{eq:Gini_uniform} with $\alpha\to\alpha_{\rm II}$, and \eqref{eq:Gini_III}, respectively. In all cases the agreement is excellent, confirming the validity of the Fokker--Planck description. The corresponding shape parameters $\alpha$ extracted from the stationary variance also match the analytical formulas (not shown).}
\label{fig:ginis_consolidados}
\end{figure*}

To verify the analytical predictions, we performed agent-based simulations of the discrete dynamics with $N=10^4$ agents, $\nu_0=1.05$, $\varepsilon=0.1$, and $10^4$ time steps, discarding the first half as transient. Figure~\ref{fig:ginis_consolidados} consolidates the comparison between the theoretical Gini coefficients and the numerical results for all three models in the moderate-parameter regime. The non-monotonic regime of Model~I, being the most dramatic finding, is treated separately in Fig.~\ref{fig:Gini_I_gamma12}.

\subsection{Aggregate growth and the trade-off}
\label{sec:model3_growth}

\begin{figure}[!htb]
\includegraphics[width=\columnwidth]{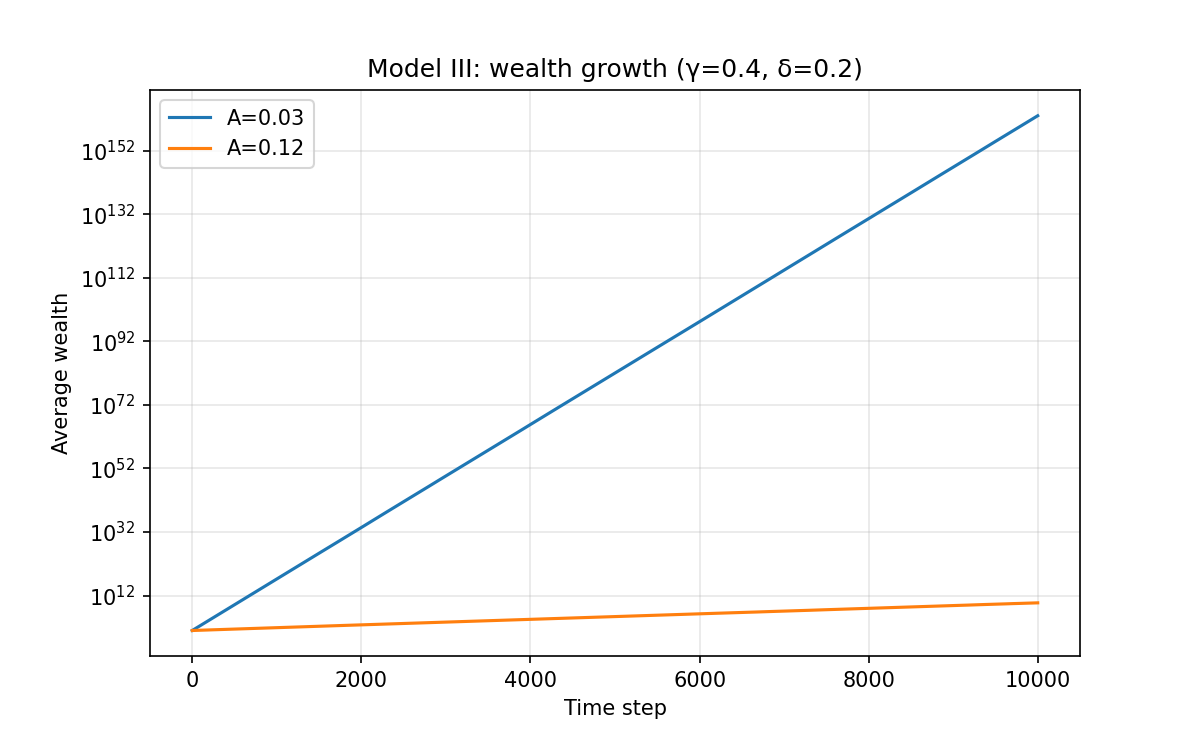}
\caption{Time evolution of average wealth for $A=0.03$ and $A=0.12$ with $\gamma=0.4$, $\delta=0.2$. The log-linear scale highlights the different growth rates; the higher tax rate yields a noticeably slower growth.}
\label{fig:crescimento_III}
\end{figure}

\begin{figure}[!htb]
\includegraphics[width=\columnwidth]{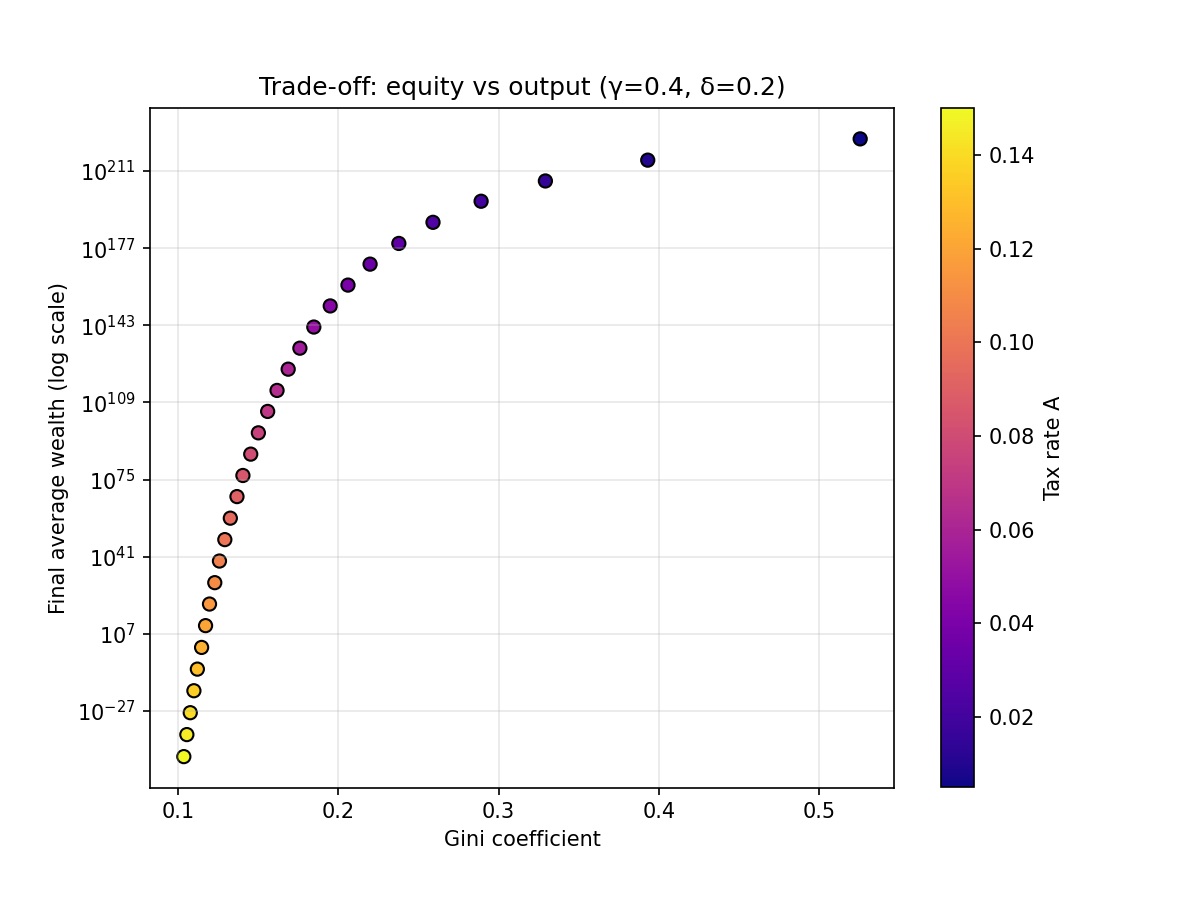}
\caption{Trade-off between equity and aggregate output in the combined model (Model~III) with fixed $\gamma=0.4$ and $\delta=0.2$. Each point corresponds to a different tax rate $A$ (indicated by the color scale), showing the stationary Gini coefficient versus the final average wealth after $T=10^4$ time steps. The monotonic decrease of wealth with decreasing inequality illustrates the efficiency cost of redistribution and gives rise to an interior optimal tax rate when both objectives are combined in a social welfare function.}
\label{fig:tradeoff}
\end{figure}

Figure~\ref{fig:crescimento_III} illustrates the growth of average wealth for two tax rates in the combined model with fixed $\gamma=0.4$ and $\delta=0.2$. The slower growth at higher $A$ is evident. The direct trade-off between equity and output is visualized in Fig.~\ref{fig:tradeoff}, where each point represents a stationary state for a different $A$. The monotonic decrease of final wealth with decreasing Gini coefficient encapsulates the efficiency cost of redistribution.

\subsection{Optimal taxation in the combined model}
\label{sec:model3_optimal}

The welfare function $\mathcal{W}_\epsilon(A; \gamma, \delta)$ is obtained by substituting Eqs.\ \eqref{eq:alpha_III} and \eqref{eq:growth_III} into the Atkinson formula. For any given $(\gamma, \delta)$ and inequality aversion $\epsilon > 0$, the optimal tax rate $A^*$ is the value that maximizes $\mathcal{W}_\epsilon$ over $A \in (0,1)$.

The dependence of $A^*$ on the parameters reveals an interesting interplay. When both $\gamma$ and $\delta$ are small, the optimal rate is relatively high, because redistribution is effective and the growth penalty is mild. As $\gamma$ increases, $A^*$ drops, reflecting the larger productivity cost of taxation. Similarly, a larger $\delta$ reduces the optimal rate because a greater share of the tax revenue is lost. In the extreme limits $\gamma \to \nu_0/A$ or $\delta \to 1$, the growth rate of the economy becomes so compromised that the welfare-maximizing tax rate approaches zero. The non-monotonic regime of Model~I further constrains the optimal rate: beyond the minimum of the Gini, both equity and output deteriorate, making the case for moderation even stronger.

\begin{figure*}[!htb]
\includegraphics[width=\textwidth]{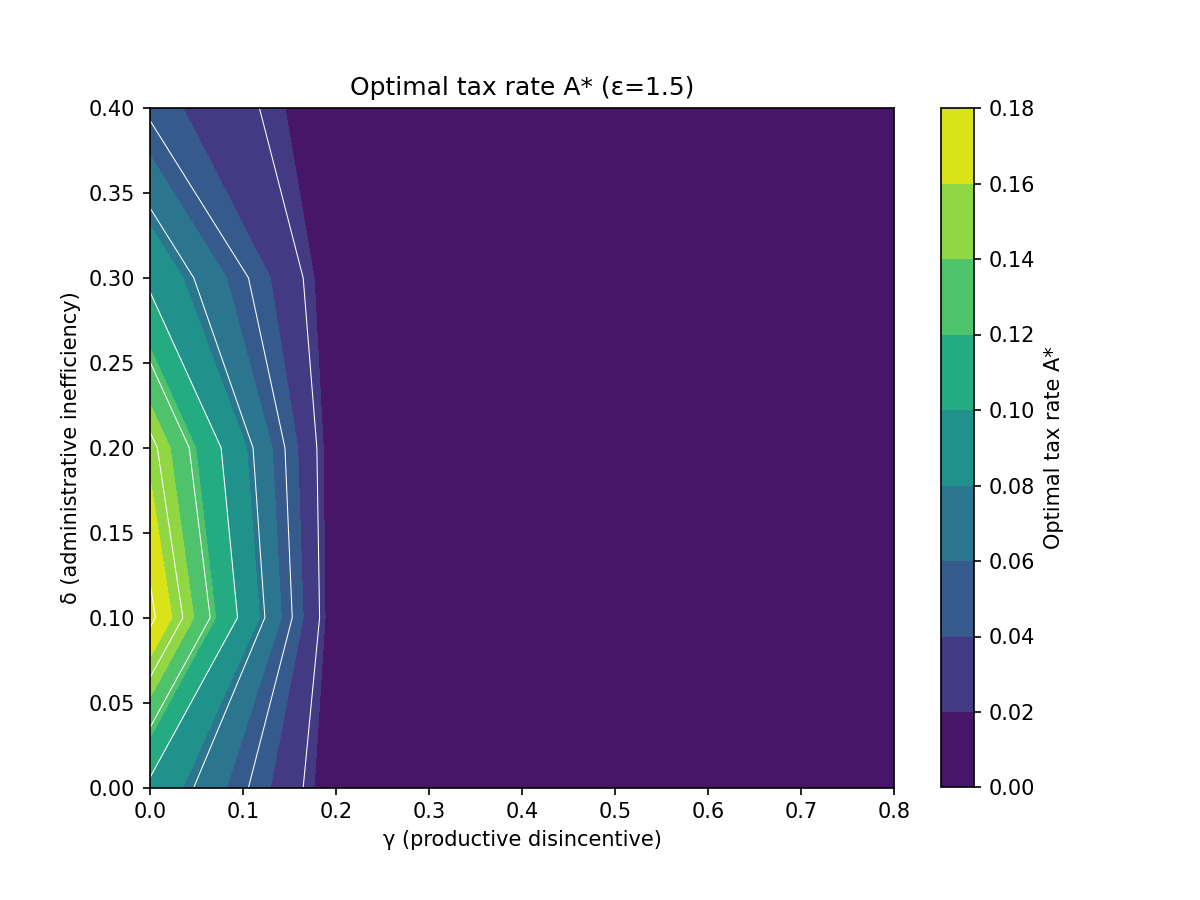}
\caption{Optimal tax rate $A^*$ that maximizes the Atkinson welfare index with inequality aversion $\epsilon=1.5$, as a function of $\gamma$ and $\delta$. The color map and contour lines show how the optimal rate decreases when either the productive disincentive or the administrative inefficiency becomes stronger. Simulation details as in previous figures.}
\label{fig:mapa_Astar}
\end{figure*}

A feature that emerges clearly from the combined model is the non-additive interaction between the two channels. Because $\alpha_{\rm III}$ contains the product $(\nu_0 - \gamma A)^2 (1-\delta)$, the effect of increasing $\gamma$ on the optimal rate depends on the value of $\delta$, and vice versa. For a fixed $\epsilon$, one can trace a locus in the $(\gamma,\delta)$ plane along which $A^*$ is constant. These iso-optimal-rate curves are not straight lines; they bend in a way that reflects the multiplicative nature of the two costs. A society with a moderately inefficient bureaucracy (say, $\delta = 0.2$) and a moderately elastic growth response (say, $\gamma = 0.3$) may have the same optimal tax rate as another with a highly efficient bureaucracy ($\delta = 0.05$) but a very elastic growth response ($\gamma = 0.6$). The precise shape of these loci depends on $\epsilon$, and in general higher inequality aversion compresses the region in which $A^*$ is significantly above zero. The resulting optimal tax rates are displayed in Fig.~\ref{fig:mapa_Astar}.

Agent-based simulations of the combined model, performed with the same protocol as before, confirm the analytical predictions. The stationary histograms of $y$ match the inverse-gamma density with $\alpha = \alpha_{\rm III}(A)$ over the entire $(\gamma,\delta)$ parameter space explored. The Gini coefficients obtained from the simulated populations agree with Eq.\ \eqref{eq:Gini_III}, and the growth of mean wealth follows Eq.\ \eqref{eq:growth_III}. The combined model thus provides a unified and analytically tractable framework for studying the double cost of redistribution.

\section{Discussion}
\label{sec:discussion}

The results presented in the previous sections demonstrate that the existence of a welfare-maximizing tax rate is a robust consequence of any mechanism that couples the aggregate growth rate to the fiscal policy. In the original model of Ref.\ \cite{Barros2026}, the growth rate $\nu$ is exogenous, and the only effect of increasing $A$ is to reduce inequality. In that setting, a planner who cares about both output and equity would always choose the maximum possible tax rate (limited only by the requirement that $\alpha > 1$ for finite variance), because there is no cost to redistribution. Once $\nu$ is allowed to depend on $A$ (either directly, as in Model~I, or indirectly through a leakage, as in Model~II), a trade-off appears and an interior optimum emerges.

The most dramatic finding of our analysis, however, is the inequality Laffer curve uncovered in Model~I. When the productive disincentive is sufficiently strong, the Gini coefficient reaches a minimum at a finite tax rate and increases thereafter. This means that beyond a certain point, taxation aggravates the very inequality it is meant to cure. In the extreme limit, the stationary state collapses to the condensed phase, restoring the oligarchic transition. This result serves as a cautionary tale: fiscal policies that ignore their impact on the productive capacity of the economy can produce perverse distributional outcomes.

This finding resonates with a broader literature in public economics that emphasizes the efficiency costs of taxation \cite{Okun1975,Mankiw2009,Saez2001}. The standard result is that the deadweight loss of taxation grows approximately with the square of the tax rate, implying that the marginal cost of redistribution increases as the tax rate rises. Our models capture a similar idea in a highly stylized but analytically transparent manner. The linear dependence of $\nu$ on $A$ in Model~I is the simplest possible assumption; a more realistic convex cost would only strengthen the interior optimum and the non-monotonic regime. Likewise, the constant leakage fraction $\delta$ could be generalized to a function $\delta(A)$ that increases with the tax rate, reflecting the idea that higher taxes invite more avoidance and evasion \cite{Tanchev2021}. The classical results on optimal capital taxation by Judd and Chamley, which established the long-run optimality of a zero capital tax \cite{Judd1985,Chamley1986}, provide a complementary perspective: while their mechanism relies on the compounding of intertemporal distortions, ours highlights the amplification of idiosyncratic risk as the source of inefficiency. In a similar spirit, the foundational work of Mirrlees on optimal income taxation \cite{Mirrlees1971} and of Stiglitz on the distributional effects of taxation \cite{Stiglitz1969} underscores the importance of incentive compatibility constraints, which in our setting are captured in reduced form by the parameters $\gamma$ and $\delta$.

It is instructive to compare the qualitative severity of the two cost channels. While both $\gamma$ and $\delta$ reduce the equalizing power of taxation, they do so in fundamentally different ways. The inefficiency parameter $\delta$ merely scales the effective tax rate by a constant factor $(1-\delta)$, leaving the monotonic relationship between $A$ and the Gini coefficient intact. The disincentive parameter $\gamma$, by contrast, enters nonlinearly through the factor $(\nu_0 - \gamma A)^2$ and simultaneously increases the effective noise amplitude $D(A)$. This dual effect---weakening the restoring drift while amplifying the multiplicative fluctuations---is what enables the non-monotonic regime and the eventual collapse to condensation. In economic terms, corruption and administrative waste reduce the amount of revenue that reaches the intended beneficiaries, but they do not make the underlying economy more volatile. A tax that severely depresses the return on productive activity, however, not only slows growth but also magnifies the relative importance of idiosyncratic shocks, thereby pushing the system toward oligarchy. This asymmetry suggests that, within the logic of the model, productive disincentive is a more dangerous enemy of equity than bureaucratic inefficiency.

Historical experience offers striking qualitative support for this asymmetry. Sweden's welfare‑state expansion in the 1970s and 1980s pushed marginal tax rates on labour and capital to levels that severely compressed after‑tax returns on productive investment \cite{Lindbeck1997}. The result was not greater equality but economic stagnation and a reversal of the previously declining trend in wealth concentration, as the tax system penalised the middle class's ability to accumulate while offering loopholes to the very wealthy. At the extreme, Venezuela's economic collapse since the early 2000s illustrates the limit $\gamma A \to \nu_0$: expropriations, price controls and institutional instability destroyed the expected return on capital, leading to a concentration of remaining wealth in the hands of a small elite connected to political power, even as official Gini measures briefly improved before deteriorating sharply \cite{Hausmann2014}. In contrast, countries that experienced significant corruption yet preserved a reasonably attractive environment for private investment---Italy during its postwar growth, or Brazil in the early 2000s---saw inequality trends that, while persistent, did not explode \cite{Mauro1995}. The combination of high corruption and low returns, as in Zaire under Mobutu, produced the worst outcome: stagnant growth and extreme wealth concentration \cite{Ndikumana2011}. These episodes, though anecdotal, resonate with the model's central prediction: fiscal policies that undermine the productive capacity of the economy can defeat their own redistributive purpose, while bureaucratic inefficiency, however costly, is less likely to destabilise the entire distributional equilibrium.

The converse implication is equally relevant. If a country finds itself on the right‑hand side of the inequality Laffer curve, where both equity and output deteriorate with further taxation, the model predicts that a tax cut raises the long‑run growth rate and, simultaneously, reduces wealth concentration. Under such circumstances, the usual trade‑off between efficiency and equity dissolves. Historical episodes are broadly consistent with this pattern, although they inevitably involve multiple confounding factors. Sweden's 1990–91 tax reform, which reduced the top marginal income tax rate from over 80\% to about 50\%, was followed by economic recovery and a stabilisation of inequality \cite{Agell1996}. The United States experienced a prolonged expansion after the marginal rate cuts of 1981 and 1986 \cite{Feldstein1995}. Ireland's reduction of the corporate tax rate from 40\% to 12.5\% coincided with a period of exceptionally rapid growth \cite{Honohan2002}. Russia's 2001 flat‑tax reform was followed by rising compliance and strong GDP growth \cite{Ivanova2005}. Conversely, the Kansas tax‑cut experiment of 2012, which sharply reduced rates without offsetting spending cuts, did not produce the predicted growth surge \cite{Gale2017}, illustrating that the specific fiscal context matters. None of these episodes constitutes a controlled test of the model; they merely suggest that the qualitative mechanism identified here — in which high marginal rates can, under certain conditions, harm both equity and efficiency — is not a theoretical curiosity but a plausible description of real‑world fiscal dynamics.

A limitation of the present analysis is its reliance on the uniform-redistribution benchmark to obtain closed-form solutions. As shown in Ref.\ \cite{Barros2026}, nonuniform redistribution kernels (such as progressive transfers or conditional cash transfer programs) can reduce inequality more effectively than the uniform scheme for the same tax rate. It would be of considerable interest to extend the trade-off analysis, and in particular the search for inequality Laffer curves, to those kernels. The stationary densities for the continuous and two-level kernels derived in Ref.\ \cite{Barros2026} are more complex than the inverse-gamma, involving confluent hypergeometric functions, but they remain amenable to numerical evaluation. One could then compute the Gini coefficient and the Atkinson welfare index as functions of $A$, $\gamma$, and $\delta$, and determine how the optimal tax rate shifts when redistribution is more sharply targeted. Intuition suggests that targeted transfers would allow a lower optimal tax rate to achieve the same level of welfare, but the precise quantitative effect remains to be explored.

Another direction for future work is to endogenize the parameters $\gamma$ and $\delta$ by introducing agent-level decisions. In Model~I, the parameter $\gamma$ is a reduced-form representation of a behavioral response. One could, for instance, allow agents to choose the fraction of their wealth to invest in a risky asset whose return is taxed, while the remainder is held in a tax-exempt safe asset \cite{Bassetti2011}. The optimal portfolio share would then depend on the tax rate, and the aggregate growth rate $\nu$ would emerge as an equilibrium outcome of these individual decisions. Such a model would connect the phenomenological parameter $\gamma$ to deeper microeconomic primitives. Similarly, $\delta$ could be derived from a model of tax enforcement and compliance \cite{Tanchev2021,Allingham1972}, where the probability of detection and the penalty for evasion determine the effective leakage.

A broader conceptual issue deserves mention. The econophysics literature on wealth distribution, including the present work, has predominantly adopted the Gini coefficient as the metric of success for fiscal policy. This focus on relative inequality, while analytically convenient and historically well-grounded \cite{Atkinson1970,Sen1997}, can obscure the distinction between reducing inequality and reducing poverty. A society with a high Gini coefficient but rapidly rising wealth for all agents, including the poorest, may be preferable, by some welfare criteria, to a perfectly equal society with stagnant or declining total wealth. In our framework, the trade-off between equity and output partially addresses this concern, because the Atkinson index explicitly weighs both dimensions. However, the index still penalizes relative dispersion rather than absolute deprivation. An extension of the model that incorporates a poverty line, perhaps defined as a fraction of the mean wealth, and that evaluates policies by their ability to lift agents above that threshold, would be a valuable complement to the present analysis. Such an approach might reveal that the optimal tax rate is lower than the one that minimizes the Gini, particularly in economies where the growth penalty of taxation is large.

The empirical calibration of the model parameters also merits attention. Cross-country panel data on tax rates, GDP growth, and Gini coefficients \cite{Piketty2014,Milanovic2016} could be used to estimate $\gamma$ and $\delta$ within the framework of the combined model. The structural equations \eqref{eq:alpha_III} and \eqref{eq:growth_III} provide a direct link between observables and the model parameters, and one could in principle perform a nonlinear least-squares fit to recover $\nu_0$, $\gamma$, $\delta$, and $\varepsilon$. The resulting estimates would not only test the empirical relevance of the model but also offer a quantitative guide to policy: given a country's estimated $\gamma$ and $\delta$, what is its welfare-maximizing tax rate for a given inequality aversion $\epsilon$? Additionally, identifying whether real economies operate in the monotonic or non-monotonic regime would have profound implications for fiscal policy design.

\section{Conclusions}
\label{sec:conclusion}

We have extended the Fokker--Planck description of the stochastic multiplicative wealth model with taxation and redistribution to include two realistic costs of fiscal policy: a productive disincentive, modeled by a tax-dependent average return $\nu = \nu_0 - \gamma A$, and an administrative inefficiency, modeled by a leakage fraction $\delta$ of the tax revenue. Both modifications preserve the analytical tractability of the original model in the uniform-redistribution benchmark: the stationary density of normalized wealth remains an inverse-gamma law, but its shape parameter acquires a richer dependence on the tax rate.

The key consequences are twofold. First, we have demonstrated the existence of an inequality Laffer curve: when the disincentive parameter $\gamma$ is sufficiently large, the stationary Gini coefficient reaches a minimum at a finite tax rate and then increases, leading to a collapse of the stationary state in the extreme limit. This shows that excessive taxation can aggravate wealth concentration and restore the oligarchic transition, a cautionary result for policy design. Second, both mechanisms generate a genuine trade-off between equity and aggregate output, giving rise to a welfare-maximizing tax rate $A^*$ for any positive degree of inequality aversion. We have derived closed-form expressions for $A^*$ in terms of the model parameters and have shown how it responds to changes in the disincentive strength $\gamma$ and the leakage fraction $\delta$. The combined model reveals that the two cost channels interact in a nonlinear fashion, so that the optimal policy cannot be deduced from the separate analyses alone.

Our results indicate that the qualitative existence of a welfare-maximizing tax rate is robust: it arises whenever the average growth rate of the economy is made sensitive to the fiscal policy, irrespective of the precise microeconomic mechanism that produces this sensitivity. This finding places the original model of Ref.\ \cite{Barros2026} as a special limiting case ($\gamma=\delta=0$) in which the growth rate is artificially insulated from policy and the trade-off disappears. By moving beyond that limit, and by revealing the possibility of perverse distributional effects at high tax rates, the present work contributes to a more complete econophysical understanding of the costs and benefits of redistributive taxation.

\begin{acknowledgments}
The simulation code and data that support the findings of this study are openly available at \url{https://doi.org/10.5281/zenodo.21365456}. The author acknowledges fruitful discussions with colleagues in the econophysics community.
\end{acknowledgments}

\end{document}